# Dissociative Single and Double Ionization of Pyridine

Sitanath Mondal[1], Brendan Wouterlood[1], Gustavo A. Garcia[2], Laurent Nahon[2], Frank Stienkemeier[1] and Sebastian Hartweg[1,*]

[1]Institute of Physics, University of Freiburg, Hermann-Herder-Straße 3a, 79104 Freiburg, Germany

[2]Synchrotron Soleil, L'Orme des Merisiers, Départementale 128, 91190 Saint-Aubin, France

*sebastian.hartweg@physik.uni-freiburg.de

**Abstract:** Dissociative ionization processes of simple heterocyclic molecules like pyridine are relevant for an understanding of radiation damage processes in biological material that occur naturally in complex condensed environments. Pyridine can thereby be considered a simple analogue of nucleobases and related ring structures are included in many important biomolecules. We present here a detailed study of dissociative single-photon single and double ionization processes using double imaging photoelectron photoion coincidence spectroscopy, supported by quantum chemical calculations. In the case of single ionization we correlate previously described cationic states to their corresponding ionic dissociation products observed at a photon energy of 23 eV, providing additional information beyond previously reported ion appearance energies. For the case of double ionization by 36 eV photons the analysis of electron-ion-ion triple coincidences provides detailed information on the onsets of various dissociative double ionization pathways, often only different by the locations of single hydrogen atoms. The detailed understanding of dissociative single and double ionization of pyridine is a prerequisite for future studies addressing radiation damage processes of such molecules in complex environments.

**Introduction**

Pyridine is among the simplest heterocyclic aromatic hydrocarbon molecules, and is used in the chemical industry as a solvent as well as a reactant[1]. Many biologically-relevant chemical compounds contain pyridine ring structures, as well as related structures including heterocycles containing more than one nitrogen atom[2]. An important example for the latter is the pyrimidine molecule that can be considered a direct precursor of the nucleobases thymine, cytosine and uracil. For this reason, pyridine is well suited as a proxy or simpler analogue for these biomolecular building blocks in studies of radiation damage processes, where specifically different ionization processes and the subsequent fragmentation and reaction dynamics are of interest. Such processes, typically occurring in aqueous environments, first require a firm understanding of the processes and dynamics occurring in isolated molecules. Additional interest in heterocyclic aromatic hydrocarbons comes from the realm of astrochemistry. Nitrogen heterocycles have been repeatedly reported to be present in carbon rich meteorites[3], but have not been directly observed in the interstellar medium, despite the significant dipole moments of simple heterocyclic molecules like pyridine and pyrrole. Understandably, there remain many open questions regarding the mechanism and reactions leading to the formation of molecules like pyridine in space[4,5], as well as to its destruction and processing by ionizing radiation and highly energetic particles.

The importance of pyridine and related molecules for a wide variety of chemical disciplines, have in the past motivated many experimental[6–11] and theoretical[10,12–15] photoelectron spectroscopic studies, elucidating the electronic structure of pyridine and its cation. Additional information has been obtained from electron impact studies, using electron energy loss techniques to obtain insights into cationic and neutral excited states beyond those that are optically accessible[16–18]. Electron impact mass spectrometry studies have also provided crucial information on dissociative ionization processes[18–20]. High mass resolution[19] as well as additional IR spectroscopic investigation of the fragments[20] have been instrumental in distinguishing possible fragments of identical mass and assigning them to certain fragments. Comparison of these studies to corresponding photoionization mass spectrometry studies[21] show good agreement.

We report here a valence shell photoelectron photoion coincidence study of pyridine allowing to correlate the population of cationic electronic states to the corresponding cationic fragmentation pathways, beyond the comparison of cationic appearance energies. Furthermore, we show that the ability to analyse electron-ion-ion triple coincidence events reliably isolates signals arising from double ionization processes, that contribute significantly to the ion yields above an excitation energy of 25 eV. Based on this distinction we provide a comparison of the cationic and dicationic dissociation processes and their respective energetic thresholds.

**Methods**

The presented experiment was performed at the DESIRS beamline[22] of the synchrotron facility Soleil in St. Aubin, France. A sample of pyridine (97% pure) was purchased from Sigma Aldrich and used without further purification. The sample was evaporated in a temperature-controlled oven, mixed with Helium carrier gas and expanded through a 50 µm nozzle into the SAPHIRS molecular beam chamber[23,24]. To produce a cluster free molecular beam the oven and nozzle were kept at 35 and 40°C, respectively and a helium pressure of 2 bar was used. The molecular beam was skimmed twice before entering the interaction region of the DELICIOUS III photoelectron photoion coincidence spectrometer[23,24]. In the interaction region of the spectrometer, pyridine was ionized by monochromatic linearly polarized XUV radiation (23 eV and 36 eV) from the monochromatized branch of the DESIRS beamline, and electrons and ions were detected in coincidence by the two designated imaging detectors of the DELICIOUS III spectrometer. The ion optics were operated in ion velocity focusing mode, to allow the retrieval of 3D momentum information of all detected photoions. The obtained photoelectron velocity map images (VMI) were filtered for coincident ion mass-to-charge ratio and photoelectron spectra and photoelectron angular distributions were reconstructed using the pBasex algorithm[25]. Prior to reconstruction, the electron signals were binned into images of varying resolution depending on the total electron count in a certain mass channel. For the study of double ionization processes the obtained data were filtered on electron-ion-ion triple coincidences. This filtering approach allows retrieving information on the mass to charge ratios of ionic fragments produced after double ionization, their ion kinetic energy release information, as well as the photoelectron kinetic energy distribution of electrons produced in these double ionization processes. Note that due to the dead time of the electron detector and signal processing electronics, individual detection of electron pairs created in double ionization events is not possible. The resulting photoelectron image is the sum of all electrons detected from double ionization events[26].

To provide deeper understanding and help in the assignments of cationic fragments we have additionally performed quantum chemical calculations. Specifically, we have used density functional theory with the B3LYP functional[27,28] with def2-TVZPPD basis set[29,30] as implemented in the ORCA program system[31–34]. To obtain adiabatic dissociative ionization and double ionization energies, we have optimized the geometries of all charged and neutral fragments, and summed their respective energies, and subtracted the optimized energy of the neutral pyridine molecule.

## Results

The mass spectrum of pyridine obtained at photon energies of 23 and 36 eV are displayed in Figure 1. At 23 eV the observed ion signals agree with previous studies[21], and we follow previous assignments for the individual ionic fragments[19] (see Table 1). At 23 eV we used the imaging capabilities of the ion detector to discard signals arising from the ionization of background gases to select events arising exclusively from the cold molecular beam. Due to the higher ion kinetic energies created, this separation based on the ion image is no longer unambiguously possible at 36 eV. The photon energy of 36 eV is above the vertical double ionization energy of 25.5 eV[35]. It is thus not immediately clear which fraction of the detected ion signals arise from dissociative single or double ionization, except for the observation of ion signals of m/z 39.5 and 38.5 which we can assign unambiguously to the intact pyridine dication $C_5H_5N^{2+}$ and the dication after loss of a neutral $H_2$ molecule $C_5H_3N^{2+}$. We will address the double ionization specifically and in detail below, after discussing the mass selected photoelectron spectra of the single ionization process in the next section.

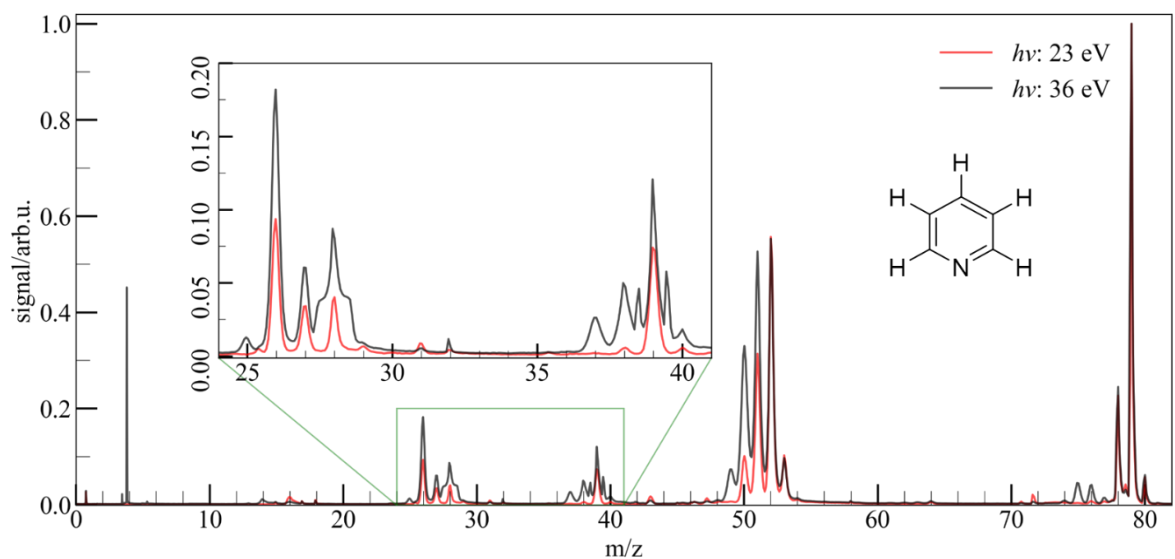

Figure 1: Ion time-of-flight mass spectra of pyridine obtained at 23 and 36 eV

## Single ionization

Figure 2 summarizes the photoelectron data obtained at 23 eV. Panel a) compares a photoelectron spectrum obtained previously at 80 eV photon energy[9] to the total photoelectron spectrum obtained in this study. Vertical lines and numbers indicate electron binding energies of previously assigned molecular orbitals (numbers 1 to 12 corresponding to orbitals HOMO to HOMO-11), where we follow the assignment and symmetry labels of Smialek et al[9]. Despite the lower resolution of our photoelectron data, the agreement between the two photoelectron spectra can be clearly observed. Variations in the relative signal intensities between 14 and 18 eV binding energy are attributed to differences in the partial cross sections due to the different photon energies used. Similar differences were also observed by Smialek et al.[9] in the comparison of threshold photoelectron spectra (TPES) with their data obtained at 80 eV photon energy. Relative intensities of the various photoelectron bands obtained at lower photon energies[11] agree more closely to our data.

The true power of PEPICO spectroscopy is that photoelectron spectra are recorded in a mass-selected fashion, allowing to clearly correlate certain dissociation pathways to the electronic states of the produced cations. We show the mass-selected photoelectron spectra in panels b)-f) of Figure 2.

The photoelectron spectrum recorded in coincidence with m/z 79 contains bands corresponding to ionization from the HOMO to HOMO-3 bands (labelled by numbers 1-4)[9]. Ionization from these four orbitals does not induce any dissociation reaction leading to fragmentation and produces stable pyridine cations. Note that we cannot detect bond dissociation reactions that lead to ring opening but not fragmentation of the molecule. Upon removal of an electron from the HOMO-4 orbital (label 5) several cationic dissociation channels open simultaneously, while a part of the cations may still remain intact. These lowest energy dissociation reactions are the loss of a hydrogen atom to form $C_5H_4N^+$ ions of m/z 78, the formation of a neutral $C_2H_2$ molecule and $C_3H_3N^+$ (m/z 53) and the formation of neutral HCN and $C_4H_4^+$ ions (m/z 52). In this assignment of the observed ion signals we follow previous work[19,20]. Our quantum chemical calculations confirm that for m/z 52 this dissociation pathway form the most stable products of the given ion masses. For m/z 53 the alternative $C_4H_5^+$ + CN was assigned by Vall-Illosera et al.[21] and would be very similar according to our calculations (Table 1). However, contributions of $C_4H_5^+$ to m/z 53 were ruled out in a high-resolution measurement using electron impact ionization[19] and seems thus less likely. These three dissociation channels remain open in parallel over an ionization energy range spanning the HOMO-4 ($1b_1$) to HOMO-9 ($8a_1$), while the branching ratios of the different channels vary slightly over this energy range. Another dissociation channel overlapping significantly with the three pathways above is the formation of $C_3H_3^+$ ions (m/z 39) and neutral $C_2H_2N$ that seemingly has a slightly higher threshold energy but covers a similar range. Note that, due to the limited energy resolution of our data and the expected vibrational envelops of the photoelectron bands, the assignments to certain electronic states are not unambiguous.

The formation of $C_4H_3^+$ ions (m/z 51) proceeds after ionization from the HOMO-9 to HOMO-11 ($8a_1$, $7a_1$ and $4b_2$) orbitals. The alternative $C_3HN+$ was not observed after electron impact[19] and would be energetically higher lying according to our calculations (Table 1). The formation of the $C_4H_2^+$ ion (m/z 50) can occur upon ionization from the HOMO-9 orbital and orbitals lying even lower than HOMO-11, while ionization from HOMO-10 and HOMO-11 does not result in ions of m/z 50 (see figure 2 c). Jiao et al[19]. argue that $C_4H_3^+$ can be either formed by elimination of neutral HCN from $C_5H_4N^+$ (m/z 78) or by loss a H atom from $C_4H_4^+$. Both possible pathways are indistinguishable in our asymptotic calculations and agree with our photoelectron data, showing that $C_4H_3^+$ ions are produced in an ionization energy range lying mainly above the range corresponding to the formation of $C_4H_4^+$ and $C_5H_4N^+$ ions. $C_4H_2^+$ ions can only be formed by successive H-loss from $C_4H_3^+$ or by loss of a neutral $H_2$ moiety from $C_4H_4^+$. Thus, it seems surprising that experimentally the energetically lowest appearances of $C_4H_3^+$ and $C_4H_2^+$ coincide. According to our DFT calculations, the asymptotic energy for the formation of $C_4H_2^+$ from $C_4H_4^+$ via the formation of neutral $H_2$ is 14.8 eV, which is slightly higher than the corresponding value for the formation $C_4H_3^+$ by the loss of a neutral $CH_2N$ from pyridine (13.9 eV). An alternative pathway for the formation of $C_4H_3^+$ via the loss of neutral H from $C_4H_4^+$ is calculated to have a higher adiabatic appearance energy (15.43 eV). It thus seems reasonable that at low ionization energy (15-18 eV) we observe the formation of $C_4H_2^+$ via the loss of $H_2$ from $C_4H_4^+$ in parallel to the formation of $C_4H_3^+$ and $CH_2N$. At higher ionization energy (above 18 eV) the rise in electron signal in coincidence with $C_4H_2^+$ ions in parallel to a decrease of electron signal in coincidence with $C_4H_3^+$ indicates an additional pathway for the formation of $C_4H_2^+$ via the loss of a hydrogen atom from $C_4H_3^+$, for which our DFT calculations predict a dissociative ionization energy of 18.07 eV.

Ions of m/z 28, 27 and 26 can be seen as the charged versions of the neutral fragments formed in the dissociation reactions described already above. We observe $C_2H_2^+$ (m/z 26) mostly at ionization energies above 16.5 eV, while the complementary $C_3H_3N^+$ (m/z 53) is mostly observed below this value. We therefore tentatively assign the two ions to the same fragmentation reaction, with the positive charge ending up on the two different molecular moieties. The DFT calculations, however, predict the asymptotic energy difference between these two channels to be only 0.62 eV. Close inspection of the photoelectron spectrum in coincidence to $C_2H_2^+$ ions (see supplementary Figure S1) reveals a weak onset around 14.3 eV in agreement with the DFT calculations. This observation indicates that the formation of the smaller ion $C_2H_2^+$ is unlikely due to kinetic reasons at low excess energy and becomes more efficient at higher excess energy. $HCNH^+$ (m/z 28) is likely formed similar to neutral HCN but includes an additional hydrogen rearrangement step prior to the dissociation or a proton transfer from $C_4H_4^+$ to HCN after or during dissociation. These ions are mostly observed in coincidence with photoelectrons corresponding to ionization from the HOMO-7 to HOMO-9 orbitals above ionization energies of 14 eV, comparing reasonably to our computational adiabatic value of 13.4 eV at the DFT level. Calculations for $C_2H_4^+$ corresponding to the same m/z as $HCNH^+$ yield a similar asymptotic threshold. Nevertheless, considering that $C_2H_4^+$ was not observed by Jiao et al.[19] and the substantial hydrogen rearrangement that would be necessary to form $C_2H_4^+$ ions, this alternative seems less likely. For ions of m/z 27 two possible ions have been reported to be formed from pyridine: $HCN^+$ and $C_2H_3^+$[19, 21]. While Jiao et al. observe similar contributions from both ions in their electron impact study, Vall-Illosera et al.[21] assume a predominant contribution of $HCN^+$ in their photoionization study. Our DFT calculations predict an asymptotic dissociative ionization energy of 16.5 eV for the formation of $HCN^+$, which is significantly above the observed onset of 14.5 eV. We therefore assign m/z 27 to $C_2H_3^+$ ions that might be formed by a proton transfer reaction during the formation of neutral $C_2H_2$ with a calculated asymptotic dissociative ionization energy of 14.5 eV but cannot exclude contributions of $HCN^+$ at higher ionization energies.

Table 1: Mass of fragments, their chemical formula and their corresponding binding energy. Cationic states assignments are taken from Smiałek et al.[9]

| Mass / amu | Formula | Measured ionization energy range / eV | Cationic states | Calculated adiabatic dissociative ionization energy / eV |
|---|---|---|---|---|
| 79 | $C_5H_5N^+$ | 9-13 | $11a_1, 1a_2, 2b_1, 7b_2, 1b_1$ | 8.97 |
| 78 | $C_5H_4N^+ + H$ | 12-18 | $1b_1, 10a_1, 6b_2, 5b_1, 9a_1$ | 12.52 |
| 53 | $C_3H_3N^+ + C_2H_2$ | 12-19 | $1b_1, 10a_1, 6b_2, 5b_1, 9a_1$ | 13.60 |
| | $C_4H_5^+ + CN$ | | | 13.61 |
| 52 | $C_4H_4^+ + HCN$ | 12-19 | $1b_1, 10a_1, 6b_2, 5b_1, 9a_1$ | 11.97 |
| | $C_3H_2N^+ + C_2H_3$ | | | 15.07 |
| 51 | $C_4H_3^+ + H_2CN$ | 15-22 | $5b_1, 9a_1, 8a_1, 7a_1, 4b_2$ | 13.89 |
| | $C_4H_3^+ + HCN + H$ | | | 15.43 |
| | $C_3HN^+ + C_2H_4$ | | | 14.31 |
| 50 | $C_4H_2^+ + H_2CN + H$ | 15-22 | $5b_1, 9a_1, 8a_1, 7a_1, 4b_2$ | 18.07 |
| | $C_4H_2^+ + HCN + H_2$ | | | 14.83 |
| 39 | $C_3H_3^+ + C_2H_2N$ | 13-20 | $10a_1, 6b_2, 5b_1, 9a_1, 8a_1$ | 12.26 |
| 28 | $CHNH^+ + C_4H_3$ | 14-22 | $5b_1, 9a_1, 8a_1, 7a_1, 4b_2$ | 13.46 |
| | $C_2H_4^+ + C_3NH$ | | | 13.36 |
| 27 | $C_2H_3^+ + C_3H_2N$ | 15-22 | $5b_1, 9a_1, 8a_1, 7a_1, 4b_2$ | 14.45 |
| | $HCN^+ + C_4H_4$ | | | 16.53 |
| 26 | $C_2H_2^+ + C_3H_3N$ | 16.5-22 | $5b_1, 9a_1, 8a_1, 7a_1, 4b_2$ | 14.22 |

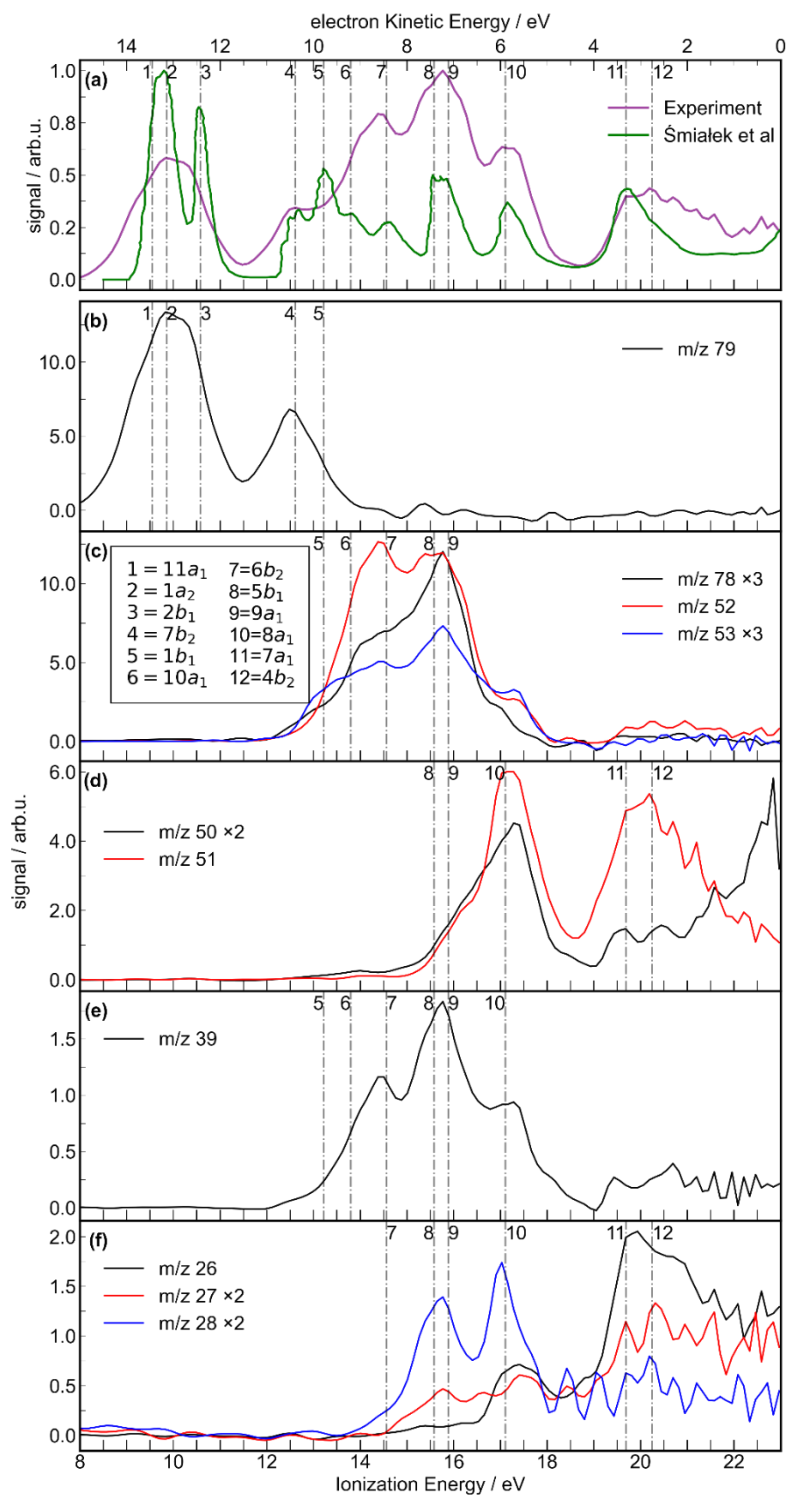

Figure 2: Photoelectron spectra of pyridine at 23 eV photon energy due to single ionization. Panel a) shows the obtained total photoelectron spectrum in comparison to the spectrum obtained by Śmiałek et al.[9] at 80 eV photon energy. The spectra in panels b)-f) show the different mass-selected photoelectron spectra, the vertical dash-dotted lines correspond to vertical transitions to cationic states assigned in reference[9].

**Double Ionization**

The only doubly-charged cations directly observable in the mass spectrum of figure 1 are $C_5H_5N^{2+}$ and $C_5H_3N^{2+}$. The loss of neutral $H_2$ molecules upon double ionization of small molecules has often been reported and recently received significant attention due to the ability of the formed $H_2$ to contribute to exotic roaming reactions[36,37]. The presence of $C_5H_4N^{2+}$ ions (m/z 39) cannot be directly confirmed in the mass spectrum. Most dissociative double ionization processes are expected to form two singly-charged rather than one doubly-charged and neutral fragments. To identify these processes, we filter the ion signals for double ionization events, i.e. events with two ions registered on the ion detector in close temporal proximity. We display this data in Figure 3 where we plot the double coincidence signals as a function of the mass-to-charge ratio of the detected ions. In this representation ion pairs created from dissociative double ionization events form anti-diagonal features, due to the correlation of their momenta along the spectrometer axis. Uncorrelated false double coincidences created from independent single ionization events form horizontal and vertical lines in this representation.

Inspection of the antidiagonal features in the region of m/z (ion1) 26-28 and m/z (ion2) 50-52 (see inset) reveals the power of this filtering approach. Features that are overlapping on both individual axes, can still be reliably separated on the two-dimensional map. We can thus immediately identify the dissociative double ionization pathways leading to the formation of pairs of singly charged ions. We list the observed ion pairs in Table 2. Many of the observed signals can be attributed to two-body fragmentation processes forming ion pairs of m/z 28-51, 27-52, 26-53 as well as 38-41, 15-64 and 14-65, where the former three are minor channels. An additional weak channel is the formation of ion pairs of m/z 1-78, attributed to the ejection of a single proton from the pyridine dication (not shown in Figure 3). The first three of these channels (m/z 26-28 with m/z 51-53) also occur accompanied by the loss of a neutral H-atom from the larger ion, producing ion pairs of m/z 28-50, 27-51 and 26-52. The only dissociation channels involving the loss of larger neutral fragments are the formation of m/z 15-37 and 14-38, along with neutral HCN molecules. It is worth noting, that ions that are observed in the mass spectrum recorded at 36 eV but not at 23 eV photon energy (Figure 1), that do not appear as ion pair signals in figure 2 can be directly assigned to products of dissociative single ionization processes becoming accessible between 23 and 36 eV. Although energetically the corresponding cationic states could undergo autoionization to form dications, the presumably ultrafast dissociation processes seem to close the autoionization channels. This includes the somewhat peculiar series of ions of m/z 77, 76, 75 and 74 ($C_5H_3N^+$, $C_5H_2N^+$, $C_5HN^+$ and $C_5N^+$) as well as m/z 49 ($C_4H^+$), 25 ($C_2H^+$). The corresponding photoelectron spectra are shown in the SI. For the series of m/z 77-74 these spectra suggest successive hydrogen or dihydrogen loss processes becoming accessible at increasing excess energy. Similarly, we assume ions of m/z 49 are produced from dihydrogen loss from $C_4H_3^+$.

To obtain a first assignment for the dicationic dissociation channels we performed DFT calculations of the possible fragments, to establish asymptotic dissociative double ionization energies. The energetically lowest fragments of a given mass are given in Table 2. For many of the observed ion pairs the DFT calculations find the same fragments as observed after single ionization. For example, the most stable ion pair of m/z 28-51 is according to our calculations $HCNH^+ + C_4H_3^+$ and for m/z 26-53 we find $C_2H_2^+ + C_3H_3N^+$ as most stable composition. The most stable ion pair m/z 27-52 on the other hand is $C_2H_3^+ + C_3H_2N^+$ where the ion of m/z 52 differs from the $C_4H_4^+$ ion observed after single ionization. The DFT calculations confirm that the energetically most stable ion pairs after hydrogen loss are formed from the above discussed ion pairs. Ion pairs of m/z 38-41 are assigned to $C_3H_2^+$ and $C_2H_3N^+$. Interestingly, the ion pair of a cyclic $C_3H_3^+$ and $C_2H_2N^+$, that we do not observe experimentally, would be considerably more stable than the latter according to the calculations. The existence of large barriers suppressing the formation of $C_3H_3^+$ seems unlikely, since we observe abundant $C_3H_3^+$ signals both at 23 eV and 36 eV. A possible explanation for the lack of ion pairs of m/z 39-40 could be that the heavier ion of m/z 40 evades detection in coincidence with m/z 39 due to the detector dead time. The minor ion pairs involving ions of m/z 14 and 15 are assigned to $CH_2^+$ and $CH_3^+$ that are initially formed alongside $C_4H_3N^+$ or $C_4H_2N^+$, where a neutral HCN molecule can still be formed to finally produce ion pairs of $CH_3^+/C_3H^+$ and $CH_2^+/C_3H_2^+$.

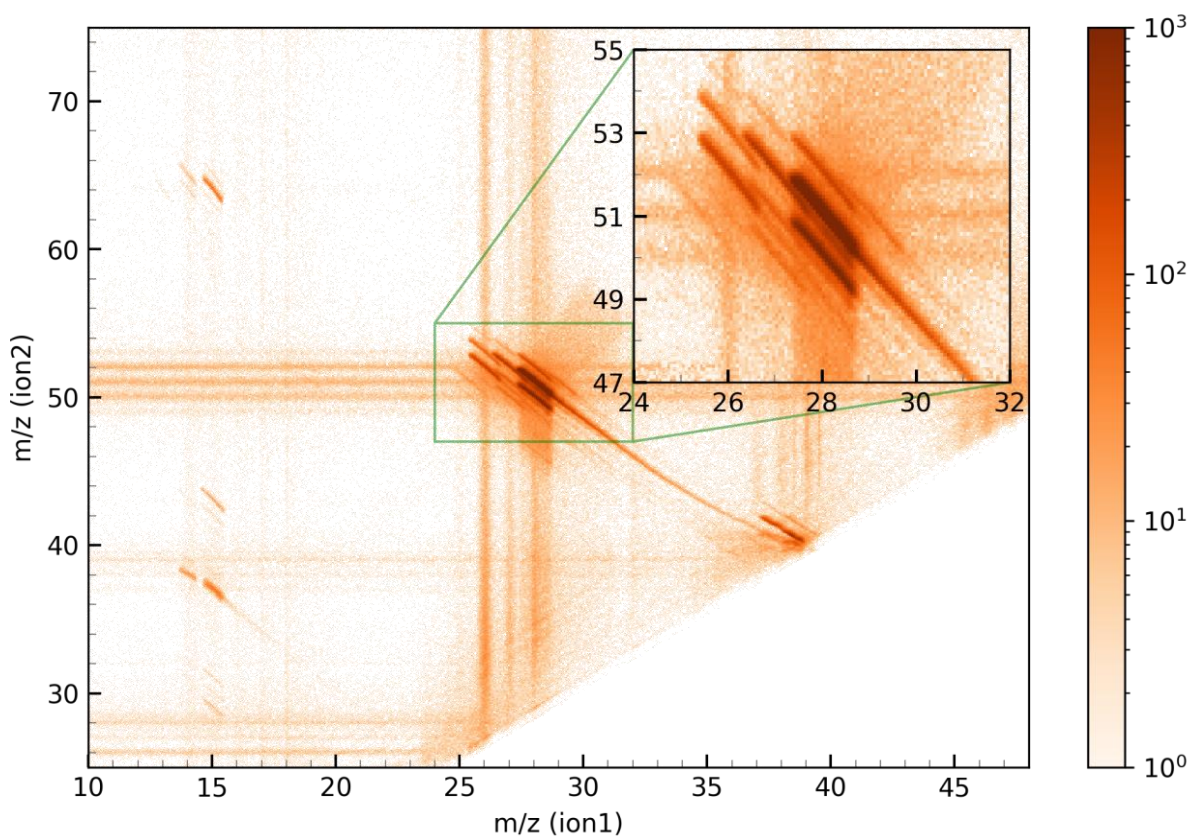

Figure 3: Ion/Ion coincidence map of pyridine at 36 eV photon energy

The ion pair signal of m/z 28-51 shows a long tail extending on the ion-ion coincidence map toward a point corresponding to 39.5-39.5. These ion pairs, characterized by two ions arriving at almost the same time, correspond to a doubly ionized pyridine molecules that undergoes fragmentation when the dication is about to leave the extraction region. The difference in ion time-of-flight of an ion pair Coulomb exploding in the field free time-of-flight region of the spectrometer will be given by the orientation-dependent momentum along the spectrometer axis. Since the energy released in the Coulomb explosion (< 5 eV) is insignificant when compared to kinetic energy of the extracted ions the time-of-flight difference of such ions will be smaller than the dead time of the ion detector and consequently those ion pairs will be detected as a single ion of m/z 39.5. The time necessary for the extraction of a pyridine dication into the field free TOF region is ~360 ns. Thus, we can conclude that there are stable dicationic electronic states, metastable electronic states that dissociate into ion pairs of m/z 28-51 on a ~100 ns timescale, and states that undergo direct dissociation into pairs of m/z 28-51 without detectable time-delay, likely on a picosecond scale.

In addition to the mass-to-charge ratio, we can analyse the ion kinetic energy release data and the photoelectron spectra detected in coincidence with the ion pairs. The kinetic energy release of a certain fragmentation channel is calculated as the sum of the kinetic energies of both individually detected ions, and the corresponding values are given in Table 2. The partitioning of the total kinetic energy is governed by momentum conservation, including also undetected neutral fragments. The average kinetic energy release for the different channels varies between 3.0 and 3.8 eV (see distributions in supplementary Table S6). The maximum distance of two atoms in the neutral pyridine molecule is about 280 pm. Assuming positive charges separated by this distance would correspond to a Coulomb repulsion energy of 5.1 eV that could be released as kinetic energy upon dissociation. While a part of this energy is undoubtedly converted to vibrational energy, it seems most plausible that upon double ionization most dissociation reactions follow a stepwise pathway first undergoing a ring opening reaction, allowing for increased charge spatial separation of the positive charges prior to final dissociation. The differences in the kinetic energy release are a strong indication for substantial differences in the mechanistic details of the various pathways.

Table 2: Ion pairs formed due to fragmentation following double ionization, their chemical formula, relative abundance and corresponding threshold double ionization energy; last lines correspond to intact dication channels.

| Ion pair m/z | Chemical formula | Yield / arb. u. | Onset pseudo electron binding energy / eV | ion kinetic energy release (KER) / eV | Calculated adiabatic (dissociative) double ionization energies / eV | Expected appearance energy / eV (KER + calculated adiabatic value) |
|---|---|---|---|---|---|---|
| 28-51 | $C_4H_3^+ + CHNH^+$ | 100 | 25.8 | 3.56 | 21.37 | 24.93 |
| 27-52 | $C_3H_2N^+ + C_2H_3^+$ | 8.43 | ~28.0 | 3.87 | 23.70 | 27.57 |
| 26-53 | $C_2H_2^+ + C_3H_3N^+$ | 1.91 | ~28.0 | 3.54 | 24.73 | 28.27 |
| 38-41 | $C_3H_2^+ + C_2H_3N^+$ | 9.0 | ~28.0 | 3.96 | 25.49 | 29.45 |
| 39-40 | $C_3H_3^+ + C_2H_2N^+$ | - | - | - | 22.47 | - |
| 14-65 | $CH_2^+ + C_4H_3N^+$ | 0.11 | n.a. | 3.12 | 26.49 | 29.55 |
| 15-64 | $CH_3^+ + C_4H_2N^+$ | 0.96 | n.a. | 3.35 | 26.32 | 29.67 |
| 28-50 | $HCNH^+ + C_4H_2^+ + H$ | 29.59 | 31.0 | 3.6 | 25.55 | 29.15 |
| 27-51 | $C_2H_3^+ + C_3HN^+ + H$ | 2.87 | 31.8 | 3.42 | 27.95 | 31.37 |
| 26-52 | $C_2H_2^+ + C_3NH_2^+ + H$ | 6.57 | 31.0 | 3.59 | 28.16 | 31.75 |
| 14-38 | $CH_2^+ + C_4H_3N^+ + HCN$ | 0.11 | >32.0 | 3.12 | 28.71 | 31.83 |
| 15-37 | $CH_3^+ + C_4H_2N^+ + HCN$ | 0.96 | 32.0 | 3.66 | 28.02 | 31.68 |
| 39.5 | $C_5H_5N^{++}$ | 33.48 | 24.5 | -- | 24.60 | 24.60 |
| 38.5 | $C_5H_3N^{++} + H_2$ | 32.93 | 28.5 | -- | 29.24 | 29.24 |

More information can be obtained from an analysis of the electron signals, detected in coincidence with the ion pairs. A thorough analysis of the double ionization spectra requires the individual detection of both electrons produced in a double ionization process. In our experiments the deadtime of the electron detector is larger than the possible differences in the flight time of two electrons produced at the same time preventing us to access a PEPEPIPICO detection scheme. We therefore detect only a single electron out of the two produced in a double ionization event. With the employed VMI detection scheme, we do not expect to observe a detection bias for faster or slower electrons. The observed electron images in coincidence with ion pair signals are therefore expected to be statistical mixtures of the produced electrons. The resulting photoelectron kinetic energy distributions are displayed in Figure 4. Panel a) shows kinetic energy distributions for two-body fragmentation processes and the double ionization processes forming intact dications. Panel b) shows the corresponding distributions for dissociation channels including the loss of neutral fragments. For completeness we also include the electron kinetic energy distributions detected in coincidence with dications. In contrast to single ionization photoelectron spectroscopy, we cannot directly link the kinetic energy of the single detected electron to the binding or double ionization energy. However, energy sharing in double ionization processes is given by a U-shaped partition curve which curvature depends on the excess available energy. In our case with a quite low excess energy, we expect a nearly flat distribution[38,39]. The fastest detected electrons therefore are created in coincidence with a electron of near zero kinetic energy and provide us with a good estimate for the threshold of various dicationic dissociation thresholds[40] allowing comparison to computational results. We will refer to electron binding energies retrieved under the assumption of a second electron with zero kinetic energy as pseudo-binding energy and note that an observed onset in pseudo-binding energy should correspond to the threshold in double ionization energy.

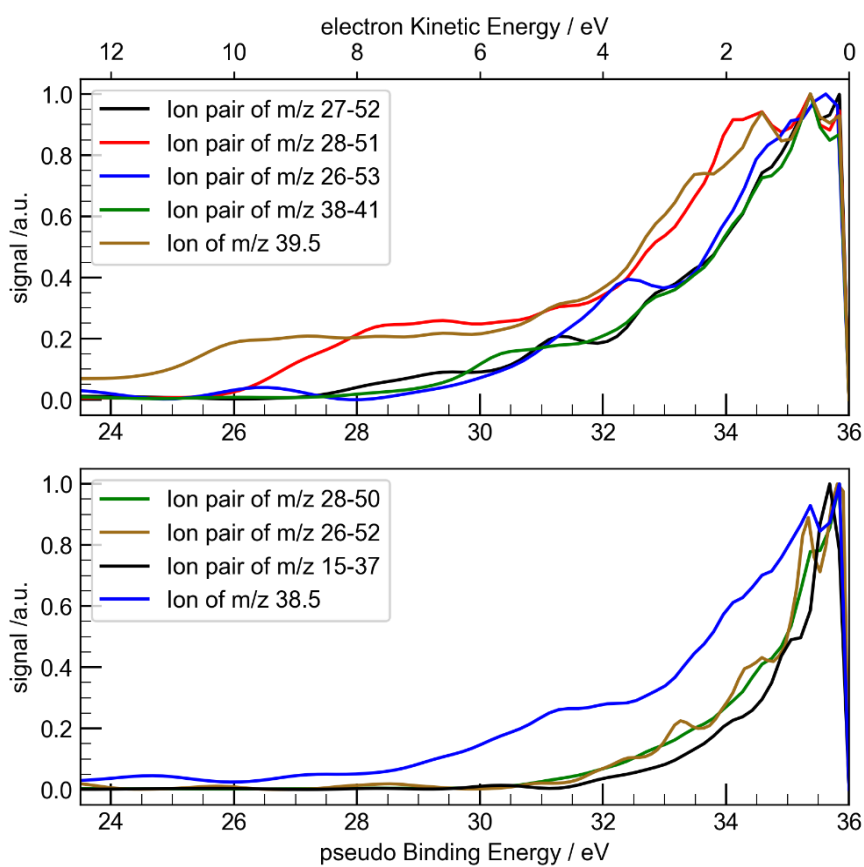

Figure 4: Mass selected photoelectron spectra from double ionization processes of pyridine at 36 eV photon energy a) without and b) with the loss of neutral fragments.

The fastest electrons (eKE<11.5 eV) are detected in coincidence with the intact dication $C_5H_5N^{2+}$ indicating that the intact dication may be attributed to the dicationic ground state. The onset of the distribution corresponds to a pseudo-binding energy of ~24.5 eV, very close to the calculated adiabatic double ionization energy of 24.6 eV. These adiabatic values agree also very well with a rule of thumb that for aromatic molecules the double ionization energy is given by 2.65 multiplied with the single ionization energy[35,41], if it is applied to the adiabatic single ionization energy of ~9.2 eV[9]. Our calculations yield a vertical double ionization energy of 25.2 eV, below the previously reported value of 25.5 eV[35]. Electrons detected in coincidence with ion pairs of m/z 28-51 are detected up to eKE<10.2 eV corresponding to a pseudo-binding energy of 25.8 eV. Similarly, the other two-body fragmentation channels at m/z 27-52, 38-41 and 26-53 are observed at higher pseudo binding energies of ~28 eV. Note that due to relatively low signal levels in these channels, we assume relatively large errors of around 1 eV on the determined thresholds. The channels including losses of neutral fragments (figure 4 b) show onsets at higher peudo-binding energy. The energetically lowest channel is the stable dication after loss of neutral $H_2$ (~28.5 eV), followed by the formation of ion pairs of m/z 26-52 and 28-50 (~31 eV) and the formation of m/z 15-37 and the loss of a neutral HCN (~32 eV).

In the comparison of our experimental data with the theoretical results a first observation is that most of the two-body dissociation channels are exothermic upon double ionization, with adiabatic dissociative double ionization thresholds below the adiabatic non-dissociative double ionization threshold. Nonetheless, the dissociative channels are only observed at higher excess energies, indicating significant energetic barriers in the dissociation pathways. To compare the calculated adiabatic dissociative double ionization energies more quantitatively with the experimentally observed onsets, we can sum the calculated values and the experimentally observed ion kinetic energy release. This approach allows a simplistic estimate of the experimentally expected threshold for a certain dissociative double ionization channel, that only neglects vibrational and rotational energy contributions of the produced fragments. The comparison between the estimated values and the experimental observations shows reasonable agreement, with deviations of less than 1 eV for most channels, despite the simplicity of the estimate. It should be noted that the expected appearance energies in Table 2 likely underestimate the true barriers and thus should in principle provide a lower limit of the appearance energies. This makes cases in which the expected appearance energies lie significantly above the experimentally observed value difficult to explain. Significant initial thermal energy in the pyridine molecules, would decrease the experimentally observe threshold artificially, but is unlikely to be a factor in our molecular beam experiment. Kinetic effects like metastability in the threshold region of a given channel would only increase the experimentally observed threshold.

**Conclusion**

In conclusion, we have presented a photoelectron photoion coincidence study of the dissociative single and double photoionization of pyridine. For the case of pyridine single ionization our data allows to connect previously observed and described cationic excited states to their respective dissociation pathway. The significant energetic overlap and lack of clear energetic ordering between various channels suggests complex dissociation pathways, possibly involving initial ring-opening reactions followed by multiple possible bond ruptures. The opening and closing of various dissociation channels observed in photoelectron spectra provides information beyond ion appearance energies and highlights that observations from electron impact and photoionization can differ significantly. The photoelectron spectra detected in coincidence with ions of m/z 51 and m/z 50 suggest different pathways involving the successive loss of neutral hydrogen atoms or the loss of neutral hydrogen molecules from larger fragments. A similar process is observed for a surprising loss of up to five hydrogen atoms for highly excited pyridine cations, above the double ionization threshold. Fragmentation products from one-photon double ionization could be reliably separated from single ionization products by filtering the obtained ion signals for electron-ion-ion triple coincidences. While some fragments are likely formed upon single and double ionization of pyridine, others like $C_3H_2N^+$, $C_3H^+$ and $C_3H_2^+$ seem to be exclusively formed upon double ionization. Particularly stable neutral fragments like HCN can be produced upon single and double ionization. Previous studies on dissociative ionization processes in pyridine, particularly employing electron ionization processes, could not distinguish between single and double ionization processes. At electron energies of ~70 eV typically used in electron ionizers for mass spectrometry, there is likely a substantial contribution of double ionization to the observed ion signals. An improved understanding of dissociative ionization processes in mass spectrometry, may therefore require inclusion of double ionization processes in their description. More detailed theoretical treatment of the various dissociation pathways, potentially including calculations of barriers and transition states as well as molecular dynamics calculations could provide deeper insights in the future, including for example more details on the dissociation pathways and the effects of vibrational energy in the final dissociation products.

**Acknowledgement**

We thank the staff of Synchrotron Soleil for smoothly running the facility under project number 20240379, and are thankful to Myriam Drissi for support in setting up the experiment. We gratefully acknowledge support by the German Research Society (DFG) in the framework of RTG 2717 the Emmy-Noether grant of S.H. (HA 10463/3-1) and through the project HA 10463/2-1.

# Supplementary Information

## for

## Dissociative Single and Double Ionization of Pyridine

Sitanath Mondal[1], Brendan Wouterlood[1], Gustavo A. Garcia[2], Laurent Nahon[2], Frank Stienkemeier[1] and Sebastian Hartweg[1,]

1. Additional Photoelectron Spectra
1.1. Photoelectron spectrum of m/z 26 at 23 eV

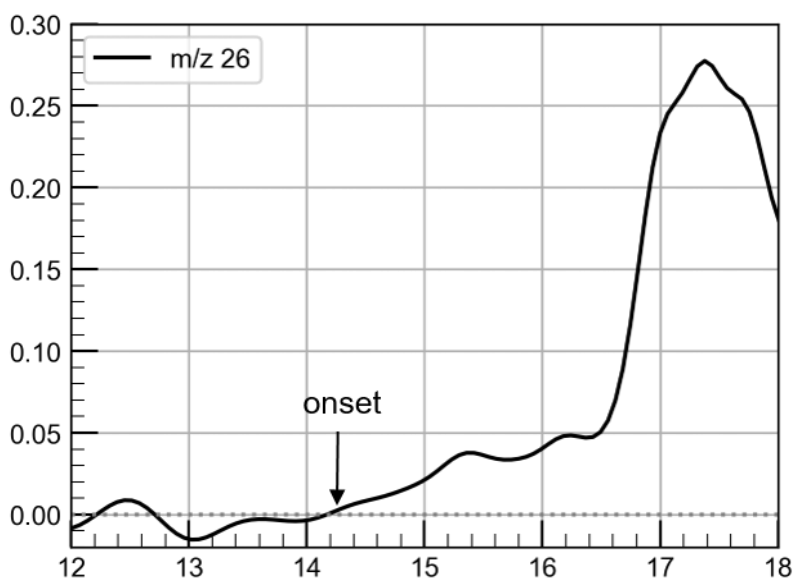

Figure S1: Onset region of photoelectron spectrum in in coincidence with m/z 26 obtained at 23 eV, compare Figure 2 in main text.

1.2. Photoelectron spectra of highly excited singly charged cations

Photoion signals that are observable only at 36 eV and not at 23 eV, but do not appear in the electron-ion-ion triple coincidence maps are assigned to fragmentation of highly excited singly-ionized pyridine molecules. Since these cations could undergo additional autoionization processes energetically, we assume that dissociation outcompetes these autoionization processes kinetically and closes the autoionization channel energetically. This mechanism seems to be the source for the production of ions of m/z 77, 76, 75, 74, 49 and 25. The former four channels show the subsequent loss of H-atoms or $H_2$ molecules from intact pyridine cations, that occur at increasing excess energy, as can be observed from the photoelectron spectra (Fig S2). Ions of m/z 49 are produced from additional H (or $H_2$) loss from $C_4H_2^+$ (or $C_4H_3^+$). Ions of m/z 25 we assign to the additional loss of H (or $H_2$) from $C_2H_2^+$ (or $C_2H_3^+$).

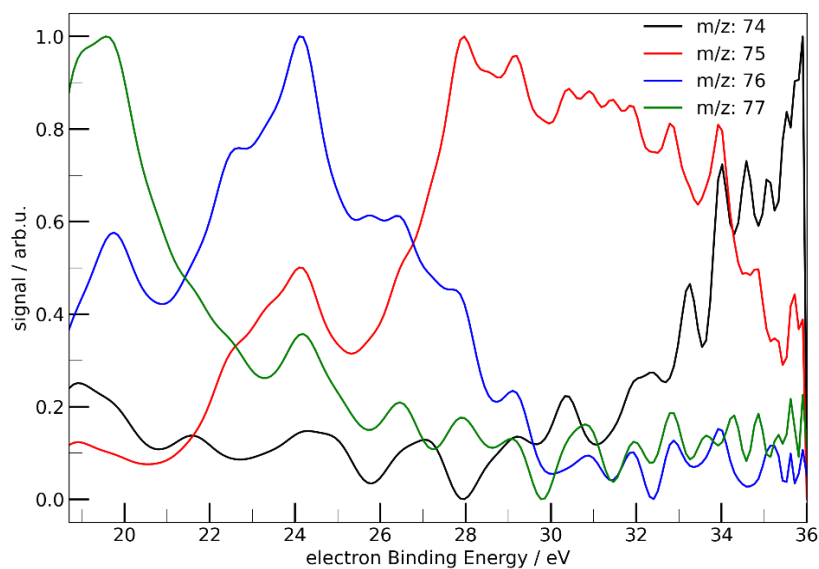

Figure S2: Photoelectron spectra obtained in coincidence with ions of m/z 77, 76, 75 and 74, corresponding to the loss of two, three four and five hydrogen atoms from a highly excited pyridine cation created at 36 eV photon energy.

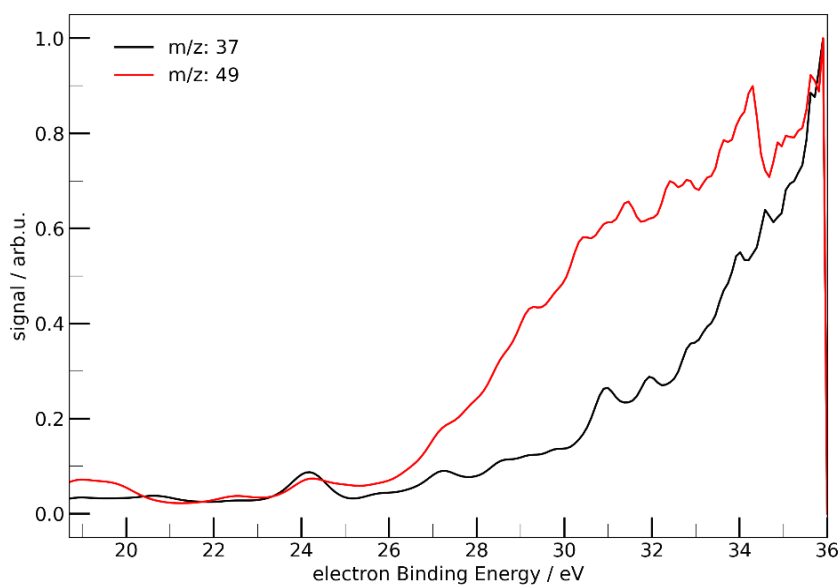

Figure S3: Photoelectron spectra obtained in coincidence with ions of m/z 49 ($C_4H^+$) and 37 ($C_3H^+$).

2. **Comparison of ionic abundances at 23 eV and 36 eV and comparison to literature.**

Table S1: Comparision of abundance of peaks between 23 eV and 36 eV from this experiment and 23 eV from this experiment and those reported by Vall-Illosera et al[21].

| Mass | 36 eV | 23 eV | From ref, 23 eV |
|---|---|---|---|
| 14 | 5.68 | 0 | |
| 15 | 2.53 | 0.47 | |
| 25 | 2.59 | 0.73 | 15 |
| 26 | 56.37 | 19.35 | 70 |
| 27 | 19.99 | 7.68 | 30 |
| 28 | 56.86 | 8.76 | 45 |
| 37 | 6.55 | 0 | |
| 38 | 10.69 | 0 | |
| 38.5 | 4.83 | 0 | |
| 39 | 20.04 | 14.65 | 17 |
| 39.5 | 4.91 | 0 | |
| 49 | 20.50 | 0 | |
| 50 | 94.44 | 22.69 | 38 |
| 51 | 113.61 | 63.17 | 47 |
| 52 | 100 | 100 | 100 |
| 53 | 21.23 | 17.91 | 15 |
| 78 | 17.52 | 21.89 | 24 |
| 79 | 63.11 | 71.84 | 63 |
| 80 | 4.65 | 4.74 | |

## 3. Quantum chemical calculations:
### 3.1.1. Optimized geometries and energies of cationic and neutral fragments

The geometry and single point energy for different mass fragments has been calculated with ORCA software using B3LYP method and def2-TZVPPD basis set. The optimized geometry of the ions are given below. In the first column the mass is given while in column 2 the chemical formula of that mass. The column 3 gives the charge and in column 4 the geometry is there. If the structures are same in non-charged fragment and the ion, only one structure is given.

Table S2: Optimized geometry for different mass fragments

| Mass | Chemical Formula | Charge | Energy | Geometry |
|---|---|---|---|---|
| 14 | N | 0 | -1485.17 | |
|  |  | 1 | -1470.61 | |
|  | $CH_2$ | 0 | 1065.1 | 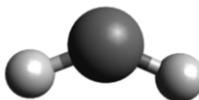 |
|  |  | 1 | 1054.77 | 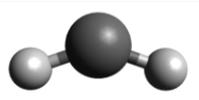 |
| 15 | NH | 0 | -1500.36 | 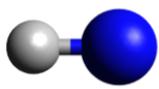 |
|  |  | 1 | 1488.96 | |
|  | $CH_3$ | 0 | -1083.77 | 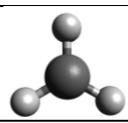 |
|  |  | 1 | -1073.96 | |
| 26 | $C_2H_2$ | 0 | -2103.84 | 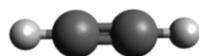 |
|  |  | 1 | -2092.68 | 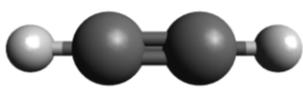 |
|  | CN | 0 | -2522.33 | 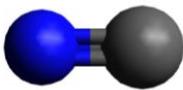 |
|  |  | 1 | -2506.07 | 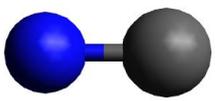 |
| 27 | $C_2H_3$ | 0 | -2119.34 | 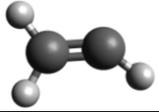 |

| | | 1 | -2110.71 | 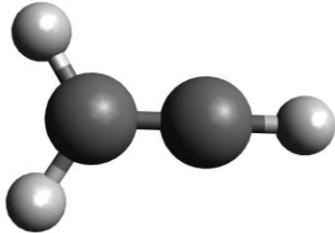 |
|---|---|---|---|---|
| | HCN | 0 | -2541.91 | 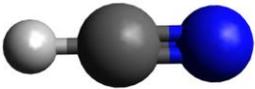 |
| | | 1 | -2528.48 | 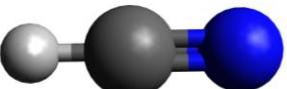 |
| 28 | C$_2$H$_4$ | 0 | -2137.93 | 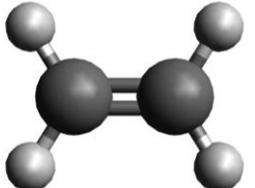 |
| | | 1 | -2127.64 | 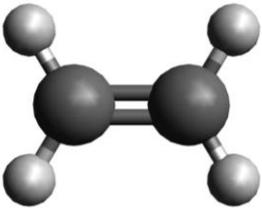 |
| | CHNH | 0 | -2557.03 | 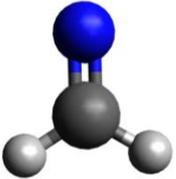 |
| | | 1 | -2549.55 | 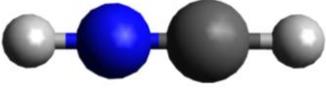 |
| 37 | C$_3$H | 0 | -3119.97 | 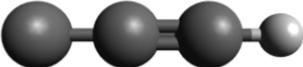 |
| | | 1 | -3110.73 | 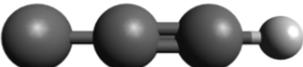 |

| 38 | C₃H₂ | 0 | -3138.24 | 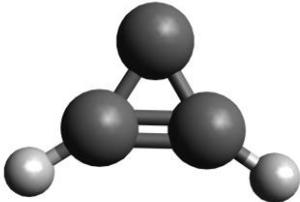 |
|---|---|---|---|---|
| | | 1 | -3129.04 | 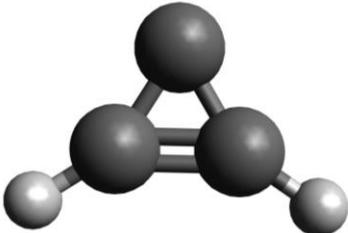 |
| | C₂N | 0 | -3557.39 | 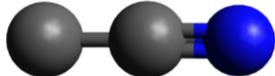 |
| | | 1 | 3546.4 | 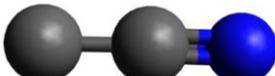 |
| 39 | C₃H₃ | 0 | -3154.16 | 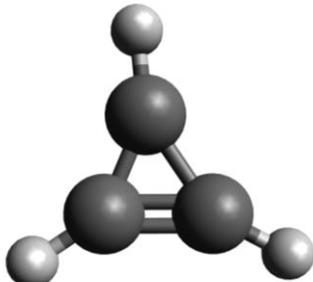 |
| | | 1 | 3148.46 | 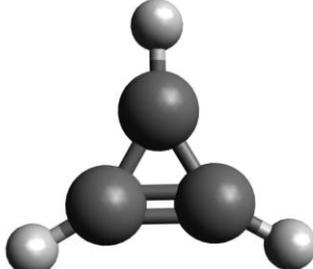 |
| | C₂NH | 0 | -3574.95 | 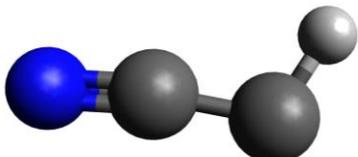 |

| | | 1 | -3565.03 | 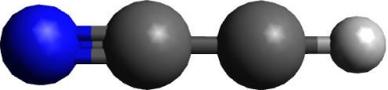 |
|---|---|---|---|---|
| 40 | C$_3$H$_4$ | 0 | -3173.66 | 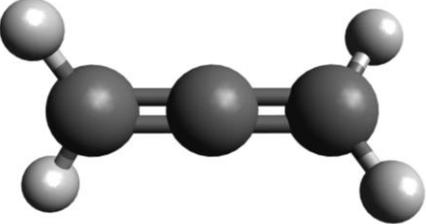 |
| | | 1 | -3164.25 | 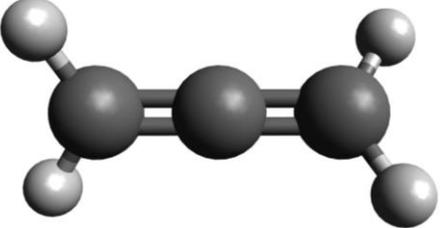 |
| | C$_2$H$_2$N | 0 | -3593.89 | 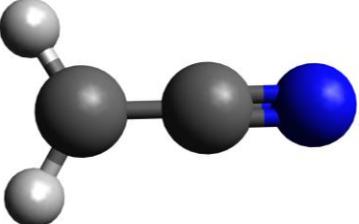 |
| | | 1 | -3583.69 | 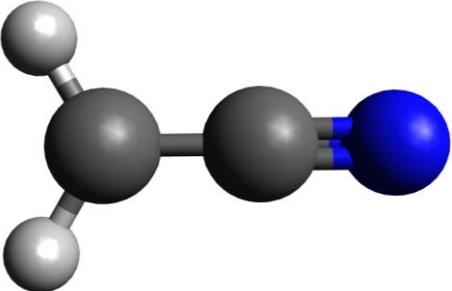 |
| 41 | C$_3$H$_5$ | 0 | -3189.98 | 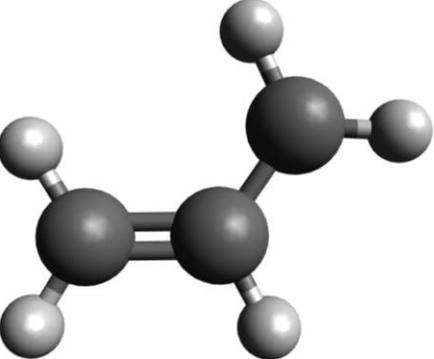 |

| | | 1 | 3182 | 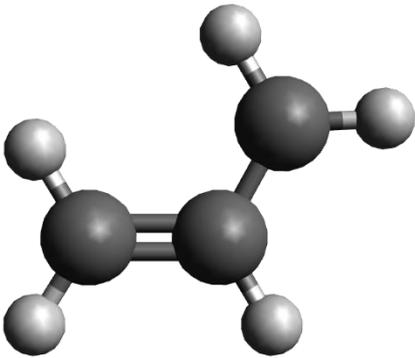 |
|---|---|---|---|---|
| | CH₃CN | 0 | -3611.79 | 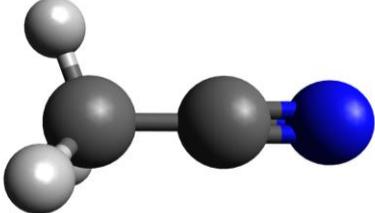 |
| | | 1 | -3599.9 | 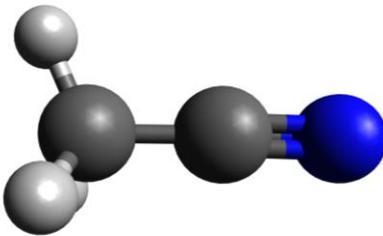 |
| 50 | C₄H₂ | 0 | -4175.7 | 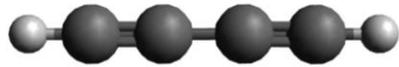 |
| | | 1 | -4165.94 | 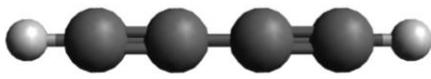 |
| | C₃N | 0 | -4593.94 | 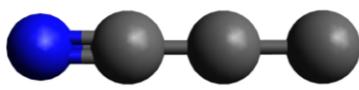 |
| | | 1 | -4581.48 | 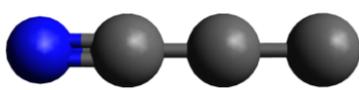 |
| 51 | C₄H₃ | 0 | -4191.6 | 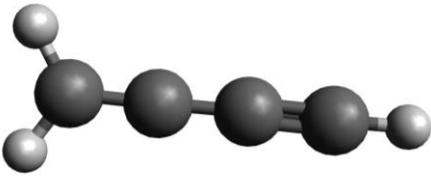 |

|    |        | 1 | -4183.7  | 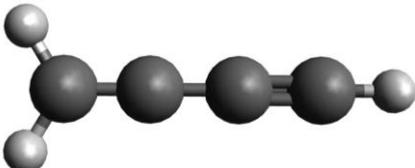 |
|    | C₃HN   | 0 | -4613.62 | 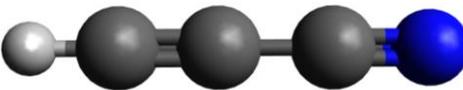 |
|    |        | 1 | -4602.38 | 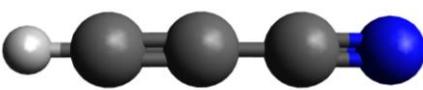 |
| 52 | C₄H₄   | 0 | -4209.6  | 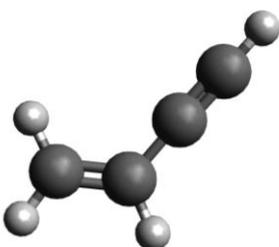 |
|    |        | 1 | -4200.74 | 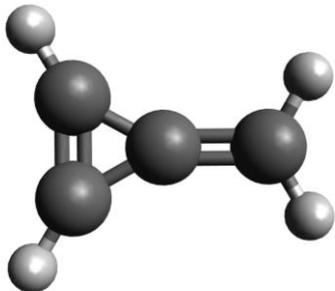 |
|    | C₃H₂N  | 0 | -4629.46 | 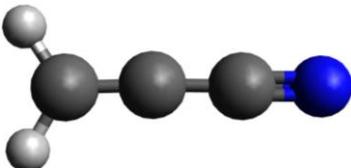 |
|    |        | 1 | -4620.2  | 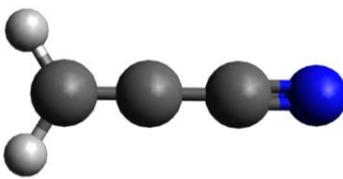 |

| 53 | C$_4$H$_5$ | 0 | -4225.46 | 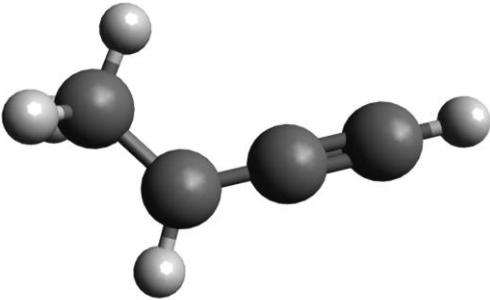 |
| --- | --- | --- | --- | --- |
|  |  | 1 | -4218.67 | 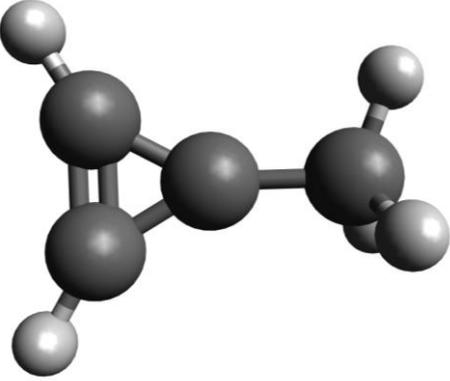 |
|  | C$_3$H$_3$N | 0 | -4647.72 | 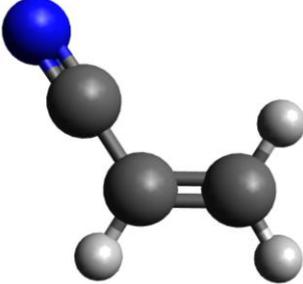 |
|  |  | 1 | -4637.2 | 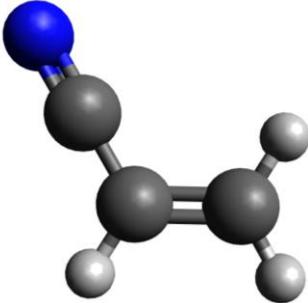 |
| 64 | C$_5$H$_4$ | 0 | -5245.22 | 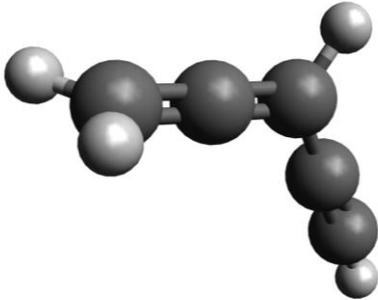 |

| | | 1 | -5235.69 | 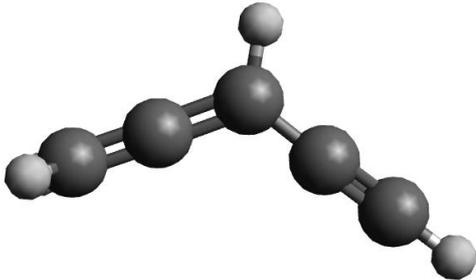 |
| --- | --- | --- | --- | --- |
| | C$_4$NH$_2$ | 0 | -5665.74 | 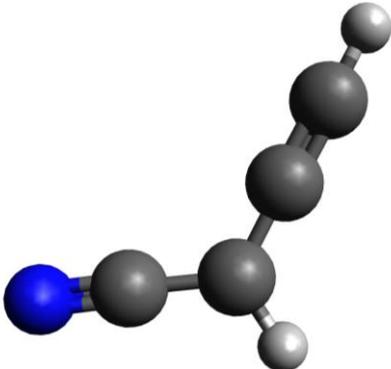 |
| | | 1 | -5654.34 | 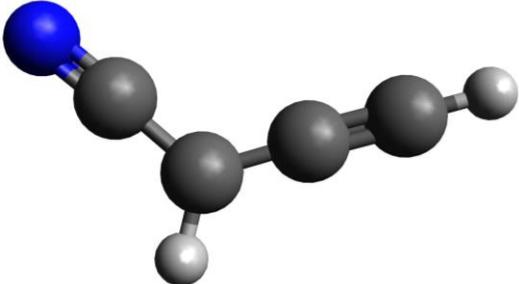 |
| 65 | C$_5$H$_5$ | 0 | -5263 | 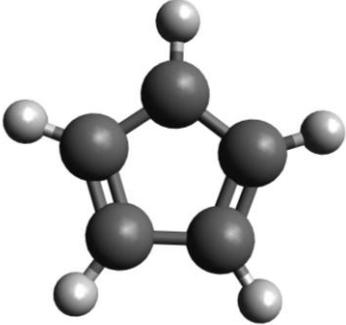 |

| | | 1 | -5254.44 | 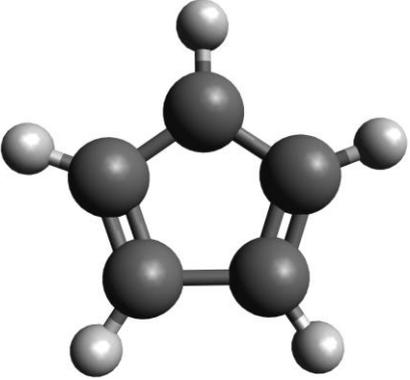 |
| --- | --- | --- | --- | --- |
| | C$_4$NH$_3$ | 0 | -5683.35 | 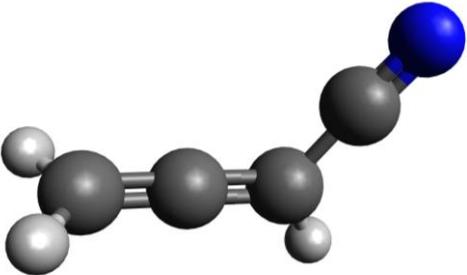 |
| | | 1 | -5673.35 | 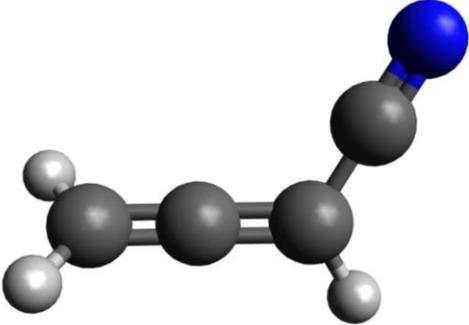 |
| 78 | C$_5$NH$_4$ | 0 | | 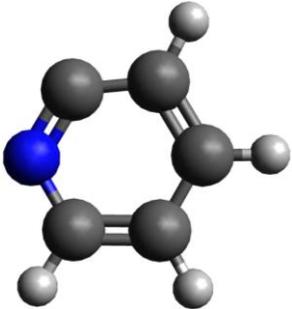 |

| | | 1 | -6728.52 | 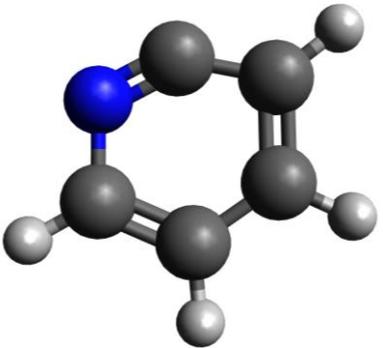 |
|---|---|---|---|---|
| 79 | C$_5$NH$_5$ | 0 | -6754.62 | 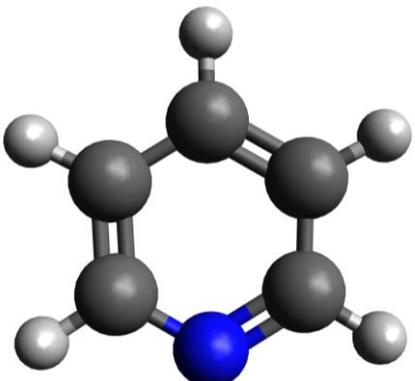 |
| | | 1 | -6745.64 | 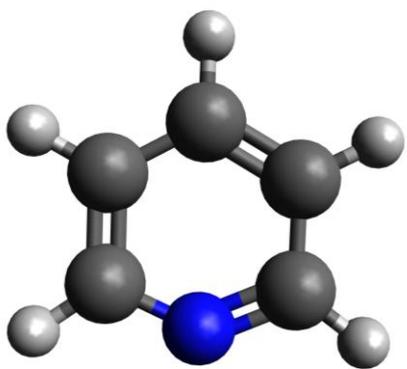 |

### 3.2. Adiabatic dissociative single ionization energies

Table S3 Calculated energies for different single ionization channels

|     | Ion | Energy(eV) |
| --- | --- | --- |
| 79 | $C_5H_5N^+$ | 8.97 |
| 78 | $C_5H_4N^+ + H$ | 12.52 |
| 65 | $C_5H_5^+ + N$ | 15.0 |
|    | $C_4H_3N^+ + CH_2$ | 16.16 |
| 64 | $C_5H_4^+ + NH^+$ | 18.57 |
|    | $C_4H_2N^+ + CH_3$ | 16.50 |
| 53 | $C_4H_5^+ + CN$ | 13.61 |
|    | $C_3NH_3^+ + C_2H_2$ | 13.57 |
| 52 | $C_4H_4^+ + HCN$ | 11.97 |
|    | $C_3H_2N^+ + C_2H_3$ | 15.07 |
| 51 | $C_4H_3^+ + CHNH$ | 13.89 |
|    | $C_3HN^+ + C_2H_4$ | 14.31 |
| 50 | $C_3N^+ + C_2H_4 + H$ | 21.63 |
|    | $C_4H_2^+ + HCNH + H$ | 18.07 |
|    | $C_4H_2^+ + HCN + H_2$ | 14.83 |
| 39 | $C_3H_3^+ + C_2H_2N$ | 12.26 |
|    | $C_2NH^+ + C_3H_4$ | 15.92 |
| 28 | $CH_2N^+ + C_4H_3$ | 13.46 |
|    | $C_2H_4^+ + C_3HN^+$ | 13.36 |
| 27 | $HCN^+ + C_4H_4$ | 16.53 |
|    | $C_2H_3^+ + C_3H_2N$ | 14.45 |
| 26 | $CN^+ + C_4H_5$ | 23.08 |
|    | $C_2H_2^+ + C_3NH_3$ | 14.22 |
| 15 | $NH^+ + C_5H_4$ | 20.43 |
|    | $CH_3^+ + C_4H_2N$ | 14.92 |
| 14 | $N^+ + C_5H_5$ | 21.00 |
|    | $CH_2^+ + C_4H_3N$ | 16.50 |

### 3.3. Adiabatic dissociative double ionization energies

Table S4: Calculated energies for different dissociative double ionization channels leading to two body fragmentation.

|       | Ion pairs                    | Energy(eV) |
|-------|------------------------------|------------|
| 52+27 | $C_4H_4^+ + HCN^+$           | 25.39      |
|       | $C_3H_2N^+ + C_2H_3^+$       | 23.70      |
| 51+28 | $C_4H_3^+ + CHNH^+$          | 21.37      |
|       | $C_3HN^+ + C_2H_4^+$         | 24.60      |
| 53+26 | $C_4H_5^+ + CN^+$            | 29.87      |
|       | $C_3NH_3^+ + C_2H_2^+$       | 24.73      |
| 14-65 | $N^+ + C_5H_5^+$             | 29.56      |
|       | $CH_2^+ + C_4H_3N^+$         | 26.49      |
| 15-64 | $NH^+ + C_5H_4^+$            | 29.97      |
|       | $CH_3^+ + C_4H_2N^+$         | 26.32      |
| 38-41 | $C_3H_2^+ + CH_3CN^+$        | 25.49      |
|       | $C_3N^+ + C_3H_5^+$          | 26.22      |
| 39-40 | $C_3H_3^+ + C_2H_2N^+$       | 22.47      |
|       | $C_2NH^+ + C_3H_4^+$         | 25.34      |

Neutral loss channels:

Table S5: Calculated energies for different dissociative double ionization channels involving the loss of neutral fragments.

|       | Ion pairs                      | Energy(eV) |
|-------|--------------------------------|------------|
| 50+28 | $C_4H_2^+ + HCNH^+ + H$        | 25.55      |
|       | $C_3N^+ + C_2H_4^+ + H$        | 31.92      |
| 51+27 | $C_4H_3^+ + HCN^+ + H$         | 28.87      |
|       | $C_3HN^+ + C_2H_3^+ + H$       | 27.95      |
| 52+26 | $C_4H_4^+ + CN^+ + H$          | 34.22      |
|       | $C_2H_2^+ + C_3NH_2^+ + H$     | 28.16      |
| 15+37 | $CH_3^+ + HCN + C_3H^+$        | 28.02      |
|       | $NH^+ + C_2H_3 + C_3H^+$       | 35.59      |

| 14-38 | $CH_2^+ + C_2H_3 + C_2N^+$ | 34.11 |
| | $N^+ + C_2H_3 + C_3H_2^+$ | 35.43 |
| | $CH_2^+ + HCN + C_3H_2^+$ | 28.71 |

Our chemical calculation gives the asymptotic energy for the channels. For any ion pair there are two possible ways. From comparing the calculated energy and the experimental KERD and photoelectron spectra, the correct channels for single and double ionization can be deduced.

4. **Ion kinetic energy release distributions of various dissociative double ionization channels.**

Table S6: Kinetic Energy Release Distributions of all observed ion pair channels.

| Ion pair | KERD | Mean KE |
|---|---|---|
| 28-51 | 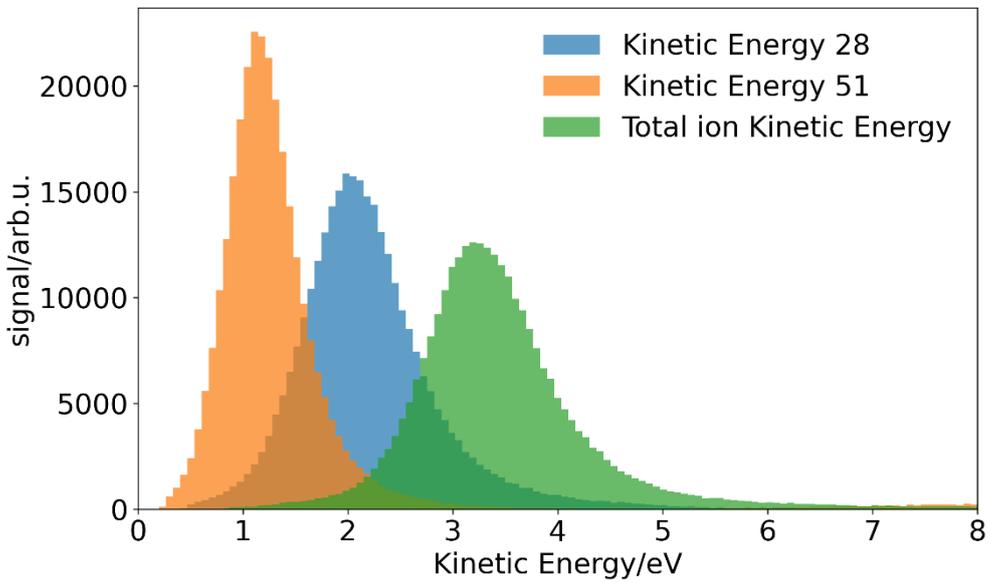 | 3.55 |
| 27-52 | 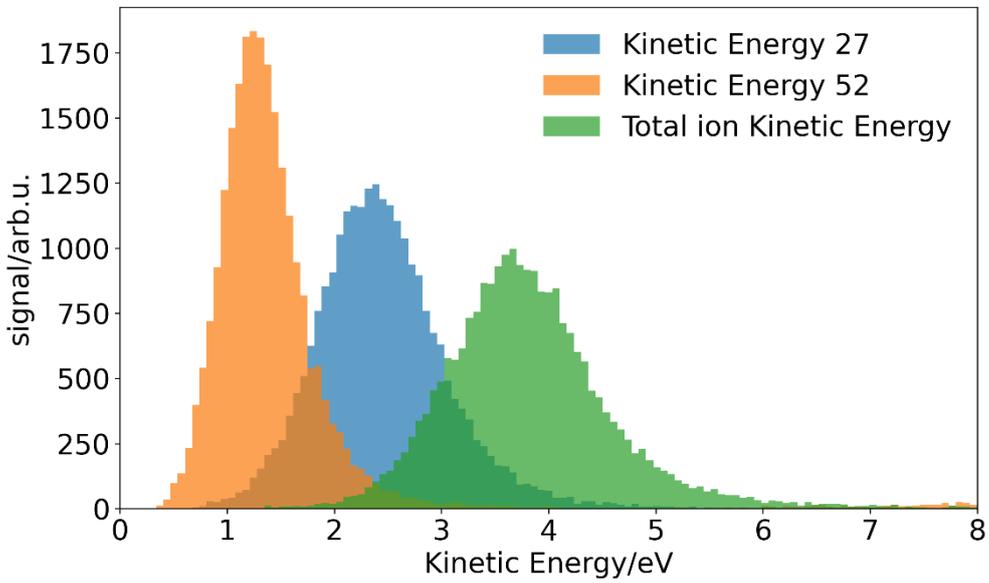 | 3.89 |
| 26-53 | 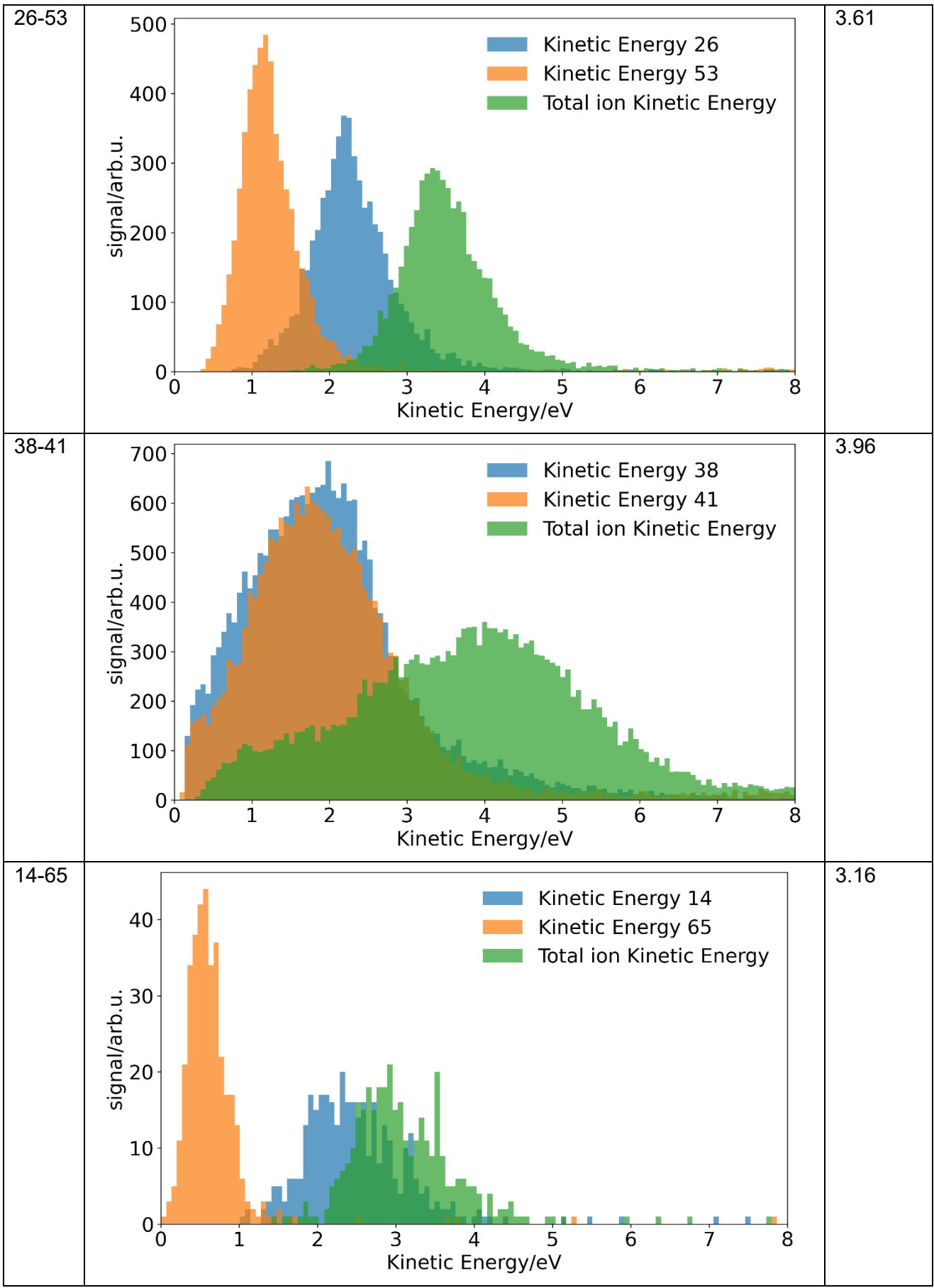 | 3.61 |
| 38-41 | | 3.96 |
| 14-65 | | 3.16 |
| 15-64 | 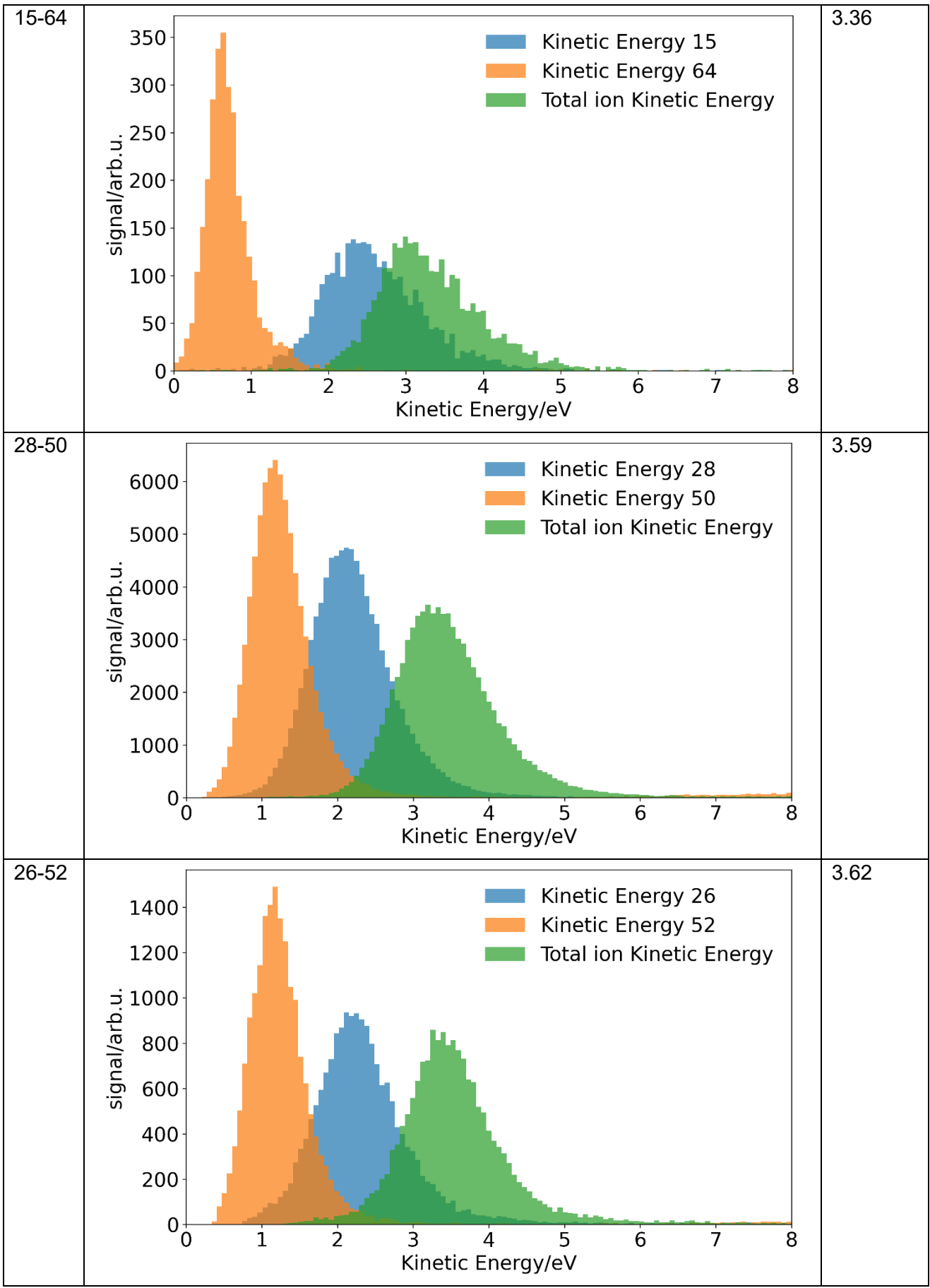 | 3.36 |
| 28-50 | | 3.59 |
| 26-52 | | 3.62 |
| 27-51 | 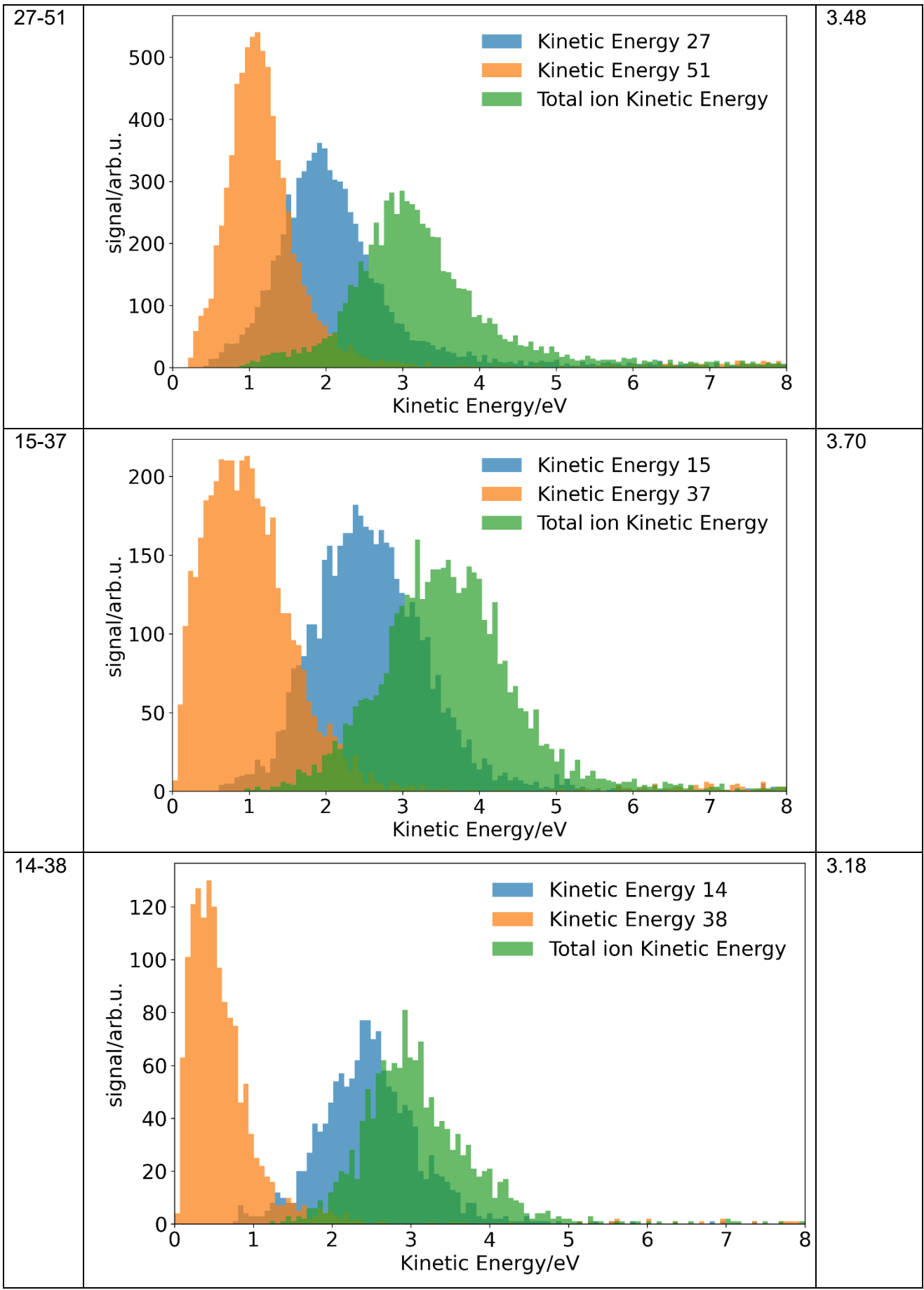 | 3.48 |
| 15-37 | | 3.70 |
| 14-38 | | 3.18 |